\documentclass[onecolumn,tighten]{aastex63}

\pdfoutput=1 
\usepackage{amsmath,amstext,longtable}
\usepackage[T1]{fontenc}
\usepackage{apjfonts} 
\usepackage[figure,figure*]{hypcap}
\shortauthors{Gramaize et al.}

\usepackage{color,amsmath,longtable}

\begin{document}

\title{Discovery of a Mid-L Dwarf Companion to the L 262-74 System}

\author[0000-0002-8960-4964]{Léopold\ Gramaize}
\affiliation{Backyard Worlds: Planet 9}

\author[0000-0002-6294-5937]{Adam C.\ Schneider}
\affiliation{United States Naval Observatory, Flagstaff Station, 10391 West Naval Observatory Road, Flagstaff, AZ 86005, USA}
\affiliation{Department of Physics and Astronomy, George Mason University, MS3F3, 4400 University Drive, Fairfax, VA 22030, USA}

\author[0000-0001-7519-1700]{Federico Marocco}
\affiliation{IPAC, Mail Code 100-22, Caltech, 1200 E. California Blvd., Pasadena, CA 91125, USA}

\author[0000-0001-6251-0573]{Jacqueline K.\ Faherty}
\affiliation{Department of Astrophysics, American Museum of Natural History, Central Park West at 79th Street, New York, NY 10034, USA}

\author[0000-0002-1125-7384]{Aaron M. Meisner}
\affiliation{NSF's National Optical-Infrared Astronomy Research Laboratory, 950 N. Cherry Ave., Tucson, AZ 85719, USA}

\author[0000-0003-4269-260X]{J.\ Davy Kirkpatrick}
\affiliation{IPAC, Mail Code 100-22, Caltech, 1200 E. California Blvd., Pasadena, CA 91125, USA}

\author[0000-0001-9482-7794]{Mark Popinchalk}
\affiliation{Department of Astrophysics, American Museum of Natural History, Central Park West at 79th Street, New York, NY 10034, USA}
\affiliation{Department of Physics, Graduate Center, City University of New York, 365 5th Ave., New York, NY 10016, USA}
\affiliation{Department of Physics and Astronomy, Hunter College, City University of New York, 695 Park Avenue, New York, NY 10065, USA}

\author[0000-0003-4083-9962]{Austin Rothermich}
\affiliation{Department of Physics, Graduate Center, City University of New York, 365 5th Ave., New York, NY 10016, USA}
\affiliation{Department of Astrophysics, American Museum of Natural History, Central Park West at 79th Street, New York, NY 10034, USA}

\author[0000-0002-2387-5489]{Marc J. Kuchner}
\affiliation{NASA Goddard Space Flight Center, Exoplanets and Stellar Astrophysics Laboratory, Code 667, Greenbelt, MD 20771, USA}

\author{The Backyard Worlds: Planet 9 Collaboration}

\begin{abstract}

We present the discovery of CWISE J151044.74$-$524923.5, a wide low-mass companion to the nearby ($\sim$24.7 pc) system L 262-74, which was identified through the Backyard Worlds: Planet 9 citizen science project. We detail the properties of the system, and we assess that this companion is a mid-L dwarf, which will need to be verified spectroscopically. With an angular separation of 74\farcs3, we estimate a projected physical separation of $\sim$1837 au from the central system.

\end{abstract}


\section{Introduction}

Backyard Worlds: Planet 9 is a citizen science project dedicated to the search of unknown nearby objects (\citealt{kuchner2017}), through the inspection of all-sky WISE images (\citealt{wright2010}). Its volunteers contributed to the discovery of several cold substellar companions to stellar systems (\citealt{faherty2020}, \citealt{faherty2021}, \citealt{schneider2021}). In an effort to identify such systems, we cross-matched a list of Gaia EDR3 sources (\citealt{gaia2021}) with the CatWISE2020 catalog (\citealt{marocco2021}), focusing the search on brown dwarf candidates. We found CWISE J151044.74$-$524923.5, a low-mass comoving source to the L 262-74 system.

\section{Properties of the central system}

L 262-74 was classified as a M2V(e) spectroscopic triple (SB3) by \cite{torres2006}. \cite{elliott2015} identified two components at an angular separation of 0\farcs47. In Gaia DR3, there are two resolved sources with similar proper motions: 5888257093214433920 and 5888257161906604800 (Table~\ref{tab:data}). For clarity reasons, we designate them as L 262-74A and L 262-74B, because L 262-74A has photometric data and the smallest astrometric error values, while L 262-74B has no available photometry. This designation may have to be reassessed in the future. From the 0\farcs873 angular separation and the respective parallaxes of these sources (Table~\ref{tab:data}), we estimate a projected separation of 21.5 to 21.6 au.

The system was part of the TESS sector 12 observation campaign (\citealt{ricker2015}). The input catalog refers to Gaia DR2 5888257093214433920 (\citealt{stassun2019}), but as the angular resolution of TESS is 21\arcsec, the system is unresolved. The light curve reveals two v-shaped $\sim$14$\%$ dips with a 17.12 day period. These dips cannot result from the orbital motion of L 262-74A and L 262-74B, because the period for two stellar objects at a separation of 21.5 au would be several years. To rule out the possibility that the eclipse signal comes from a nearby background source, we computed the photocenter of the blended detection in the individual FFIs using the {\it estimate\_centroids} function within the {\it lightkurve} python package. Since L 262-74 is the brightest source, the photocenter will be close to its position. When an eclipse happens, the system becomes fainter, so the photocenter should move slightly towards the west, where most of the background sources are. It is indeed what we observe during each transit, meaning that the signal is associated with L 262-74. Assuming that the light from L 262-74A is the dominant contribution to the light curve, we use the 14$\%$ transit depth and a 0.44R$_\odot$ radius estimate for a M2V star to assess a lower limit of 0.16R$_\odot$ for the size of the transiting companion. Finally, the Gaia DR3 re-normalized unit weight error of L 262-74A is 2.482, which could be caused by a close companion (\citealt{belokurov2020}), although such a high value could be biased (\citealt{apellaniz2021}). These hints suggest that the central system is probably triple, confirming the observation of \cite{torres2006}.

\section{Analysis of the companion}

Through visual inspection, CWISE J151044.74$-$524923.5 appears to be comoving with L 262-74. To verify this observation, we assessed the difference between the proper motion vectors (\citealt{lepine2007}) for L 262-74 (from Gaia DR3) and CWISE J151044.74$-$524923.5 (from CatWISE2020). For L 262-74A and L 262-74B, we calculate a reference value of $\Delta\mu_{AB} = 30.3 \pm 0.3$ mas yr$^{-1}$. Then, we find $\Delta\mu = 20.6 \pm 10.8$ mas yr$^{-1}$ for L 262-74A and CWISE J151044.74$-$524923.5, and $\Delta\mu = 40.2 \pm 12.3$ mas yr$^{-1}$ for L 262-74B and CWISE J151044.74$-$524923.5. Comparing these values with $\Delta\mu_{AB}$, we consider that the proper motion match between the companion and L 262-74 is similar to the one between L 262-74A and L 262-74B.

 From the 2MASS J$-K_s$ = 1.759 mag color, we estimate that the spectral type of CWISE J151044.74$-$524923.5 is L5 to L7 (\citealt{best2018}). Alternatively, the W1$-$W2 = 0.314 mag color suggests a spectral type of L4. However, the presence of a nearby potential contamination source during the NEOWISE survey epochs makes 2MASS data more reliable. Thanks to the J, H and $K_s$ absolute magnitudes inferred from the spectral type estimate, we determine a photometric distance of 17 to 30 pc, which is compatible with an association with L 262-74, whose distance is 24.6 to 24.7 pc according to the Gaia DR3 parallaxes of L 262-74A and L 262-74B.

From their positions in 2MASS, the angular separation between the companion and L 262-74 is 74\farcs6, which corresponds to a projected separation of 1844 au, assuming a distance of 24.718 pc. Proceeding similarly with the positions in CatWISE2020, we find a 74\farcs3 angular separation and a 1837 au projected separation.

Using a 52.5 mas yr$^{-1}$ matching tolerance, which corresponds to the highest possible $\Delta\mu$ value calculated earlier, we find 8\,591 Gaia DR3 sources whose proper motion match with CWISE J151044.74$-$524923.5. 173 of them are located between 17 and 30 pc. Using a projected separation of 20\,600 au as a reference (\citealt{lepine2007}, \citealt{schneider2021}), we calculate a $7.11 \times 10^{-4}$ chance alignment probability that one of these 173 sources would randomly fall within the 20\,600 au radius. Using the CoMover code (\citealt{gagne2021}), we confirm this result by determining association probabilities of 99.8\% between L 262-74A and CWISE J151044.74$-$524923.5, 95.5\% between L 262-74B and CWISE J151044.74$-$524923.5, and 94.0\% between L 262-74A and L 262-74B.

\section{Conclusions\label{sec:conclusion}}
We confirmed that L 262-74 is composed of two stellar objects, and probably of a third companion whose nature remains to be determined. We also presented the discovery of CWISE J151044.74$-$524923.5, a wide companion to this system, that we classify as a mid-L dwarf candidate, which will need to be verified spectroscopically. Using two methods, we determined consistent low chance alignment probabilities. We conclude that this companion and L 262-74 are associated, implying that the system might be quadruple. Although similar hierarchical systems like GJ 570 have been previously discovered (\citealt{burgasser2000}), only a few of them are currently known.

\section{Acknowledgements}
This work has made use of data from:
\begin{itemize}
\item the NASA NEOWISE mission, which is a joint project of Caltech/JPL and the University of Arizona.
\item the ESA {\it Gaia} mission.
\item the TESS mission. The data is accessible in MAST:\dataset[10.17909/qh1f-ww20]{http://dx.doi.org/10.17909/qh1f-ww20}.
\item Lightkurve, a Python package for Kepler and TESS data analysis (\citealt{Lightkurve 2018}).
\item the cross-match service from CDS, Strasbourg (\citealt{boch2012}).

\end{itemize}

We thank Frank Kiwy for his methodological advice.\\

{\it Software}: AstroToolBox (\citealt{kiwy2022}), CoMover (\citealt{gagne2021}), Wiseview (\citealt{caselden2018}).

\begin{deluxetable}{ccccccc}
\tabletypesize{\scriptsize}
\tablecaption{Properties of the system}\label{tab:data}
\tablehead{
\colhead{\textbf{Parameter}} &
\colhead{\textbf{L 262-74AB}}&
\colhead{\textbf{L 262-74A}} &
\colhead{\textbf{L 262-74B}} &
\colhead{\textbf{CWISE J151044.74$-$524923.5}} &
\colhead{\textbf{Unit}} &
\colhead{\textbf{Ref.}}
}
\startdata
\hline
\hline
\multicolumn{7}{c}{\textbf{Alternative Designations}} \\
\hline
\hline
2MASS& J15104047$-$5248189&...&...&J15104501$-$5249211& ...& 1 \\
Gaia DR3&...& 5888257093214433920 & 5888257161906604800& ...& ...& 2 \\
CWISE& J151040.23$-$524821.5 &...&... & J151044.74$-$524923.5&...& 3 \\
TIC&124350360&...&...&...&...&4\\
\hline
\hline
\multicolumn{7}{c}{\textbf{Astrometry}} \\
\hline
\hline
$\alpha_{2MASS}$ & 227.668633 & ... & ... & 227.687559 & deg & 1 \\
$\delta_{2MASS}$ & $-$52.80526 & ... & ... & $-$52.82254 & deg & 1 \\
$\alpha_{Gaia DR3}$ & ... & 227.6676445 & 227.6672734 & ... & deg & 2 \\
$\delta_{Gaia DR3}$ & ... & $-$52.8060496 & $-$52.8059578 & ... & deg & 2 \\
$\alpha_{CatWISE2020}$ & 227.6676556 & ... & ... & 227.6864521 & deg & 3 \\
$\delta_{CatWISE2020}$ & $-$52.8059889 & ... & ... & $-$52.8232175 & deg & 3 \\
$\mu_\alpha$ &...& $-$137.192 $\pm$ 0.036 & $-$165.654 $\pm$ 0.167 & $-$139.78 $\pm$ 7.9& mas yr$^{-1}$& 2,3 \\
$\mu_\delta$ &...& $-$169.999 $\pm$ 0.039 & $-$159.644 $\pm$ 0.174 & $-$190.46 $\pm$ 9.1& mas yr$^{-1}$& 2,3 \\
$\omega$ &...& 40.4567 $\pm$ 0.0401 & 40.6222 $\pm$ 0.1048  & ... & mas& 2 \\
$d$ &...& 24.718 $\pm$ 0.024 & 24.617 $\pm$ 0.063 & ... & pc & 2 \\
RV& 19.5 &...&... & ...& km s$^{-1}$& 5 \\
$RV_{Gaia DR3}$& ...& ... & 12.274 $\pm$ 2.48 & ...& km s$^{-1}$& 2 \\
RUWE&...&2.482 & 1.326 & ...&...&2 \\
\hline
\hline
\multicolumn{7}{c}{\textbf{Photometry}} \\
\hline
\hline
$G_{BP}$&...& 11.654 $\pm$ 0.004& ...& ...& mag& 2 \\
G&...& 10.548 $\pm$ 0.003& ...& ...& mag& 2 \\
$G_{RP}$&...& 9.226 $\pm$ 0.005& ...& ...& mag& 2 \\
J& 7.701 $\pm$ 0.021&...&...& 16.071 $\pm$ 0.12 & mag& 1 \\
H& 7.098 $\pm$ 0.042&...&...& 14.884 $\pm$ 0.073 & mag& 1 \\
$K_s$& 6.847 $\pm$ 0.023&...&...& 14.312 $\pm$ 0.076 & mag& 1 \\
W1& 6.911 $\pm$ 0.021&...&...& 13.338 $\pm$ 0.018& mag& 3 \\
W2& 6.592 $\pm$ 0.009&...&...& 13.024 $\pm$ 0.012& mag& 3 \\
W3& 6.512 $\pm$ 0.017&...&...& ...& mag& 6 \\
W4&  6.404 $\pm$ 0.059&...&...& ...& mag& 6 \\
\hline
\hline
\multicolumn{7}{c}{\textbf{Spectral type}} \\
\hline
\hline
Sp. Type&  M2&...&...& ...& ...& 5 \\
J$-K_s$ estimate&...&  ...&...& L5, L7& ...& 7 \\
W1$-$W2 estimate&...&  ...&...& L4& ...& 7 \\
\hline
\hline
\multicolumn{7}{c}{\textbf{Fundamentals\tablenotemark{a}}} \\
\hline
\hline
T$_{eff}$ &  3471 $\pm$ 157&...&...& ...& K& 4 \\
log g&  4.63 $\pm$ 0.01&...&...& ...& dex (with g in cm s$^{-2}$)& 4 \\
Radius&  0.63 $\pm$ 0.02&...&...& ...& R$_\odot$& 4 \\
Mass&  0.61 $\pm$ 0.02&...&...& ...& M$_\odot$& 4 \\
Luminosity&  0.05 $\pm$ 0.01&...&...& ...& L$_\odot$& 4 \\
\hline
\hline
\enddata
\tablecomments{Reference codes:
(1) \cite{skrutskie2006},
(2) \cite{gaia2022},
(3) \cite{marocco2021},
(4) \cite{stassun2019},
(5) \cite{torres2006},
(6) \cite{kirkpatrick2014},
(7) \cite{best2018}
}
\end{deluxetable}

\end{document}